\documentclass[twocolumn,superscriptaddress,showpacs,pra]{revtex4}
\usepackage{epsfig}
\usepackage{epstopdf}
\usepackage{delarray}
\usepackage{amsmath, amssymb}
\usepackage{bm}
\usepackage{graphicx}
\usepackage{placeins}
\usepackage{color}

\begin{document}
\title{An extended Kundu-Eckhaus equation for modeling dynamics of rogue waves in a chaotic wave-current field\footnote{This paper is to be presented at the 'Advances in Civil Engineering' conference that will be held in Istanbul,Turkey in 2016.}}

\author{Cihan Bayindir}
\email{cihan.bayindir@isikun.edu.tr}
\affiliation{Department of Civil Engineering, Isik University, Istanbul, Turkey}

\begin{abstract}
In this paper we propose an extended Kundu-Eckhaus equation (KEE) for modeling the dynamics of skewed rogue waves emerging in the vicinity of a wave blocking point due to opposing current. The equation we propose is a KEE with an additional potential term therefore the results presented in this paper can easily be generalized to study the quantum tunneling properties of the rogue waves and ultrashort (femtosecond) pulses of the KEE. In the frame of the extended KEE, we numerically show that the chaotic perturbations of the ocean current trigger the occurrence of the rogue waves on the ocean surface.  We propose and implement a split-step scheme and show that the extended KEE that we propose is unstable against random chaotic perturbations in the current profile. These perturbations transform the monochromatic wave field  into a chaotic sea state with many peaks. We numerically show that the shapes of rogue waves due to perturbations in the current profile closely follow the form of rational rogue wave solutions, especially for the central peak. We also discuss the effects of magnitude of the chaotic current perturbations on the statistics of the rogue wave occurrence.

\end{abstract}
\maketitle

\section{\label{sec:level1} Introduction}
Rogue (freak) waves are extensively studied in recent years \cite{Akhmediev2011, bayindir2016}. These studies are of crucial importance for the safety of the marine travel and offshore operations as it is important to avoid rogue waves in the open ocean to prevent damage and loss of lives. Rogue waves are also widely studied in many other fields such as Bose-Einstein condensation, optics and finance just to mention a few \cite{Wang}. Discovery of the rational rogue wave solutions for the well-known nonlinear Schr\"{o}dinger equation (NLSE) resulted in seminal studies of rogue wave dynamics, such as \cite{Akhmediev2009b}. However, the NLSE has its limitations due to the underlying assumptions and approximations used in its derivation. Therefore more general equations are derived and are being developed currently.

Rogue waves emerge due to several mechanisms in the ocean. These include nonlinear interactions, atmospheric forcing, focusing due to changes in the ocean bottom topography \cite{Kharif}. Another mechanism which triggers the rogue wave formation is the blocking thus steepening of the waves due to opposite current. In 1976, R. Smith proposed the formula
\begin{equation}
i\psi_t + \lambda \psi_{xx} +  \zeta \left|\psi \right|^2 \psi - \gamma x\left|\frac{dU}{dx} \right|\psi =0
\label{eq011}
\end{equation}
to study the nonlinear wave field in the vicinity of blocking point \cite{Smith1976}. In this formula $x, t$ denote the spatial and temporal variables, $i$ denotes the imaginary number, $\lambda, \zeta, \gamma$ are some real constants, $\psi$ is complex amplitude, $U$ is the current speed and $dU/dx$ is the current gradient calculated in the vicinity of the blocking point due to Taylor series expansion used in its derivation \cite{Kharif}. The blocking point can be found by equating the group velocity to zero, that is
\begin{equation}
C_g=\frac{d\omega}{dk}= \frac{1}{2}\sqrt{\frac{g}{k}} + U(x_o)=0
\label{eq02}
\end{equation}
where $C_g$ is the group velocity,  $x_o$ is the blocking and $g$ is the gravitational acceleration. Angular frequency, $\omega$, is given by
\begin{equation}
\omega(k)=\pm \sqrt{gk} + k U(x_o)
\label{eq03}
\end{equation}
The extended NLSE of Smith in Eq.\,(\ref{eq011}) is also used for investigating the nonlinear effects near caustics \cite{PeregrineSmith}.

In order to study the effect of quintic nonlinear and Raman-effect terms on the chaotic wave-current field and study the skewed rogue waves appear due to wave blocking in an opposing current we propose
\begin{equation}
\begin{split}
i\psi_t + \lambda \psi_{xx} + \zeta \left|\psi \right|^2 \psi + \beta^2 \left|\psi \right|^4 \psi & - 2 \beta i \left( \left|\psi \right|^2 \right)_x \psi \\
&- \gamma x\left|\frac{dU}{dx} \right|\psi =0
\label{eq04}
\end{split}
\end{equation}
where $\beta$ is a real constant and $\beta^2$ is the coefficient of the quintic nonlinear term which accounts for the effects of higher order nonlinearity \cite{Wang, dqiu}. The cancellation of the last term in this equation leads to the Kundu-Eckhaus equation. It is important to note that letting $- \gamma x\left|\frac{dU}{dx} \right|=V$, where $V$ is a potential function, the equation becomes KEE with potential function. Therefore proposed equation can be used as a model to study the quantum tunneling properties of ultrashort (femtosecond) pulses or rogue waves in the frame of the KEE. Special potentials may be used to ensure that ultrashort pulses are tunneling for intended purposes. In this study we focus on wave blocking due to opposing current in a hydrodynamic medium. A transformation of the form 
\begin{equation}
\begin{split}
\psi(x,t)= \widetilde{\psi} & \left (x+ \lambda \gamma t^2 \left|\frac{dU}{dx} \right|, t \right) \\
& .\exp{\left( -i \left[\gamma xt \left|\frac{dU}{dx} \right|+ \lambda \gamma^2 \frac{t^3}{3} \left|\frac{dU}{dx} \right|^2 \right] \right)}
\label{eq05}
\end{split}
\end{equation}
transforms the extended KEE to the KEE in the form of 
\begin{equation}
i\widetilde{\psi}_t + \lambda \widetilde{\psi}_{xx} + \zeta \left|\widetilde{\psi} \right|^2 \widetilde{\psi} + \beta^2 \left|\widetilde{\psi} \right|^4 \widetilde{\psi} - 2 \beta i \left( \left|\widetilde{\psi} \right|^2 \right)_x \widetilde{\psi} =0
\label{eq06}
\end{equation}
Some analytical periodic as well as rational solutions of the KEE exist in the literature \cite{Wang, dqiu}. For $\lambda=1, \zeta=2$ the first order rational solution, also known as the Peregrine soliton solution, of the KEE is given by
\begin{equation}
\widetilde{\psi}_1=\exp{\left[i (- \beta x+ (\beta^2+2)t) \right]} \frac{H_1+i J_1}{M_1} \exp{ \left[i \beta \frac{K_1}{M_1}  \right]}
\label{eq07}
\end{equation}
where 
\begin{equation}
\begin{split}
& H_1=-4x^2-16 \beta t x-16(\beta^2+1)t^2+3, \ \ \ J_1=16t, \\
& M_1=4 x^2+ 16 \beta t x + 16(\beta^2+1)t^2+1, \\
& K_1=4 x^3+ 16 (\beta^2+1) t^2 x + 9x+ 16\beta(x^2+1)t.
\label{eq08}
\end{split}
\end{equation}
This solution is given in \cite{Wang, dqiu} and it is basically a skewed Peregrine rogue wave of the NLSE given in \cite{Akhmediev2009b}.  Second and the higher order rational solutions of the KEE and a hierarchy of obtaining those rational solutions based on Darboux transformations are given in \cite{Wang}. As discussed in the next section a chaotic wave-current field produces only few rogue waves with a peak larger than 3, therefore we limit our rogue wave shape analysis to the first order rational solution of the KEE.

\section{\label{sec:level1}Chaotic Wave-Current Field in the Frame of Extended Kundu-Eckhaus Equation}

Although the processes described by the extended KEE seeded by the modulation instability are very complicated and hard to predict, they are still governed by a partial differential equation. Therefore for a given initial condition, they can be predicted. Therefore compared to the completely unpredictable stochastic processes, the processes studied in the frame of the extended KEE in this paper can be named as 'chaotic' processes. A similar terminology can be found in \cite{Akhmediev2014, Akhmediev2015rwsSS}.

In a recent paper we have studied the chaotic wave fields generated in the frame of the KEE \cite{bayindir2016KEE}. In that study, by calculating the probability density functions (PDFs) for various scenarios we have showed that the $\beta$ parameter controls the skewness of the wave field and smaller values of initial seed wavenumber ($k$) result in a higher probability of high amplitude waves. Our results have also demonstrated that the quintic nonlinear term in the evolution equation leads to higher probabilities of rogue wave occurrence in a chaotic wave field however the Raman-effect term in the Kundu-Eckhaus equation reduces the probability of rogue wave occurrence. Seeded by the modulation instability with a noise magnitude $0.2$, such a chaotic wave field is likely to produce many rogue waves with amplitudes around $3$, however only few rogue waves with amplitude bigger than $3$. It is natural to expect that higher the noise amplitude, higher and more rogue waves will appear in the chaotic wave field. Results for the KEE will not be reproduced here for the sake of brevity. The reader is referred to  \cite{bayindir2016KEE} for the detailed analysis of contribution of individual terms of the KEE on rogue wave formation.

We turn our attention to study the chaotic wave-current fields generated in the frame of the extended KEE proposed. In order to analyze chaotic wave-current fields in the frame of the KEE we utilize a numerical technique. We start chaotic wave-current field simulations using a constant amplitude wave and apply chaotic perturbation to the ocean current gradient. Such a state rapidly evolves into a full-scale chaotic field similar to the fields analyzed in \cite{bayindir2016, Akhmediev2009b,  Akhmediev2009a}. The chaotic wave-current field modeled by the extended KEE with this starter evolves into a wave field which exhibits many amplitude peaks and eventually some of them becoming rogue waves. In order to model such a chaotic wave-current field we use the initial conditions of
\begin{equation}
\left|\psi \right|_0=\left|\exp{i(kx)} \right|=1
\label{eq09}
\end{equation}
where $k$ denotes the initial seed plane wavenumber and chaotic perturbation in the current profile is seeded by
\begin{equation}
\frac{dU}{dx} (x,t)=\mu+\alpha [r_1+i r_2]
\label{eq10}
\end{equation}
where $i$ is the imaginary number, $\mu$ and $\alpha$ are some constants which determine the magnitude of chaotic perturbations and $r_1$, $r_2$ are uniformly distributed random number vectors admitting values in the interval of [-1,1]. It is possible to add chaotic perturbations with a characteristic length scale of $L_{pert}$, by multiplying the expression in Eq. (\ref{eq10}) with a factor of $\exp(i 2 \pi / L_{pert}x)$. Or using Fourier analysis it is possible to add perturbations with different characteristic length scales. However in present work for illustrative purposes we do not use such as scale. It is known that chaotic perturbations in the current profile impose chaotic perturbations in the ocean surface fluctuation field and the reverse of this statement is also true. Therefore we use the expression in Eq.(\ref{eq10}) continuously in a time loop to model this phenomena. The actual water surface fluctuation for wave-current field would be given by the real part of $ \left|\psi\right| \exp{[i\omega t]}$ where $\omega$ is a carrier wave frequency however we only consider the envelope $\left|\psi\right|$ in our simulations.

For the numerical solution of the KEE we propose and implement a split-step Fourier method (SSFM). In SSFM schemes spectral techniques are used to evaluate the spatial derivatives \cite{bayindir2009, Karjadi2010, Karjadi2012, bayindir2016earlyCS, bay2015d, bay2015e, trefethen}. Those spectral techniques make use of the orthogonal transforms for this purpose. The most popular choice for the periodic domain calculations is the Fourier transform where efficient fast Fourier transform (FFT) algorithms are commonly used \cite{bayindir2016, bayindir2016nature,  demiray}. SSFM performs time integration by exponential time stepping. SSFM can easily be applied to equations with first order time derivatives. A SSFM implementation usually depends on splitting the equation into two parts, the nonlinear and the linear part \cite{bayindir2015d, bayindir2015arxivchbloc, bayindir2015arxivcssfm, bay2015c, bayindir2015arxivcsmww}. For the extended KEE proposed in this study, the advance in time due to nonlinear part for $\lambda=1, \zeta=2$ can be written as
\begin{equation}
i\psi_t= -(2\left| \psi \right|^2 +  \beta^2 \left| \psi \right|^4-2i\beta [\left| \psi \right|^2]_x -x\left|{dU}/{dx} \right|   )\psi
\label{eq11}
\end{equation}
which can be exactly solved as
\begin{equation}
\tilde{\psi}(x,t_0+\Delta t)=e^{i(2\left| \psi_0 \right|^2 +  \beta^2 \left| \psi_0 \right|^4-2i\beta [\left| \psi_0 \right|^2]_x -x\left|{dU}/{dx} \right|  )\Delta t}\ \psi_0   
\label{eq12}
\end{equation}
where $\psi_0=\psi(x,t_0)$ is the initial seed plane-wave and $\Delta t$ is the time step. Using the Fourier series one can write
\begin{equation}
\begin{split}
\tilde{\psi}(x,t_0+\Delta t)=&  e^{i(2\left| \psi_0 \right|^2 +  \beta^2 \left| \psi_0 \right|^4} \\
& ^{-2i\beta F^{-1}[ikF[\left| \psi_0 \right|^2]] -x\left|{dU}/{dx} \right|  )\Delta t}\ \psi_0  
\end{split} 
\label{eq13}
\end{equation}
where $k$ is the Fourier transform parameter. Operations $F$ and $F^{-1}$ denote the forward and inverse Fourier transforms, respectively \cite{bayindir2016}. The remaining part of the extended KEE is the linear part and it can be separated as
\begin{equation}
i\psi_t=-\psi_{xx}
\label{eq14}
\end{equation}
This linear part of the extended KEE can be evaluated using the Fourier series as
 \begin{equation}
\psi(x,t_0+\Delta t)=F^{-1} \left[e^{-ik^2\Delta t}F[\tilde{\psi}(x,t_0+\Delta t) ] \right]
\label{eq15}
\end{equation}
where $k$ is the Fourier transform parameter. Therefore combining Eq. (\ref{eq13}) and Eq. (\ref{eq15}), the complete operation in a time step of the SSFM can be written as
 \begin{equation}
\begin{split}
\psi(x,t_0+\Delta t)= & F^{-1}  [e^{-ik^2\Delta t}.F[e^{i(2 | \psi_0 |^2 +  \beta^2 | \psi_0 |^4} \\  
& ^{-2i\beta F^{-1}[ikF[| \psi_0 |^2]] -x\left|{dU}/{dx} \right|  )\Delta t}\ \psi_0 ] ]
\end{split}
\label{eq16}
\end{equation}
Starting from the chaotic initial condition described above by Eq. (\ref{eq09}) and Eq. (\ref{eq10}), the numerical solution of the extended KEE is obtained for later times by the SSFM. This form of the SSFM utilizes four fast Fourier transform (FFT) operations in a time step. $N=4096=2^{12}$ spectral components are selected in order to make use of the FFTs efficiently. The time step is chosen as low as $dt=10^{-4}$ which give stable results in all runs.

In numerical runs we use a wavelength of $k=1$ for the initial plane-wave number. For this selection Eq. (\ref{eq02}) gives $U(x_o) \approx -1.57$ for $x_o=0$ to be a blocking point. For this plane-wave the angular frequency given by  Eq. (\ref{eq03}) becomes $\omega \approx 1.57$. Chaotic perturbations imposed on the current  profile rapidly transforms the monochromatic wave field into a chaotic sea state with many apparent peaks. A spatial domain of $L=[-100, 100]$ is selected for the numerical simulations.  

\begin{figure}[htb!]
\begin{center}
   \includegraphics[width=3.4in]{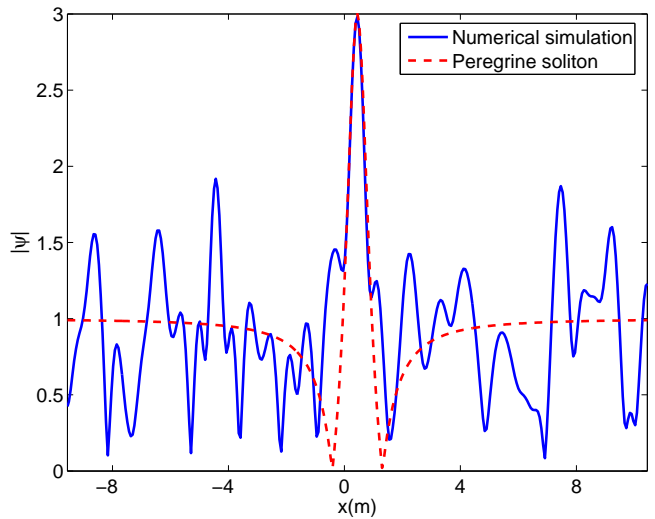}
  \end{center}
\caption{\small (Color online) Comparison of the numerical ({\color{blue} ---})  and the analytical ({\color{red} - . -}) rogue wave profiles.}
  \label{fig1}
\end{figure}

The peak value for one of the apparent rogue wave in this simulation is almost 3.00 hence it can be described by the first order rational soliton of the extended KEE. Therefore we focus on the first order rational soliton solution of the KEE obtained after application of the transform Eq. (\ref{eq05}) to extended KEE. In order to analyze the shape of the rogue wave appearing in the simulations, in the Fig.~\ref{fig1} we present its comparisons with the first order rational soliton solution of the KEE defined by Eq. (\ref{eq07}).  In this plot, the continuous blue line is taken from the numerical simulations and the exact first order rational (Peregrine) soliton is shown by the dashed red line. The central part of the peak of the rogue wave accurately follows the exact profile. The discrepancy in the tails of the peak occurs due to random smaller amplitude waves of the chaotic field that surround the peak. This result is similar to the results given in \cite{Akhmediev2009a} obtained for NLSE. Additionally it is important to note that the rogue waves in the form the rational solitons can also be described by the collision of Akhmediev breathers \cite{Akhmediev2009a}, thus the rogue wave solutions of the extended KEE can possibly be described by those collisions as well which can be the subject of a future analysis. 

\begin{figure}[htb!]
\begin{center}
   \includegraphics[width=3.4in]{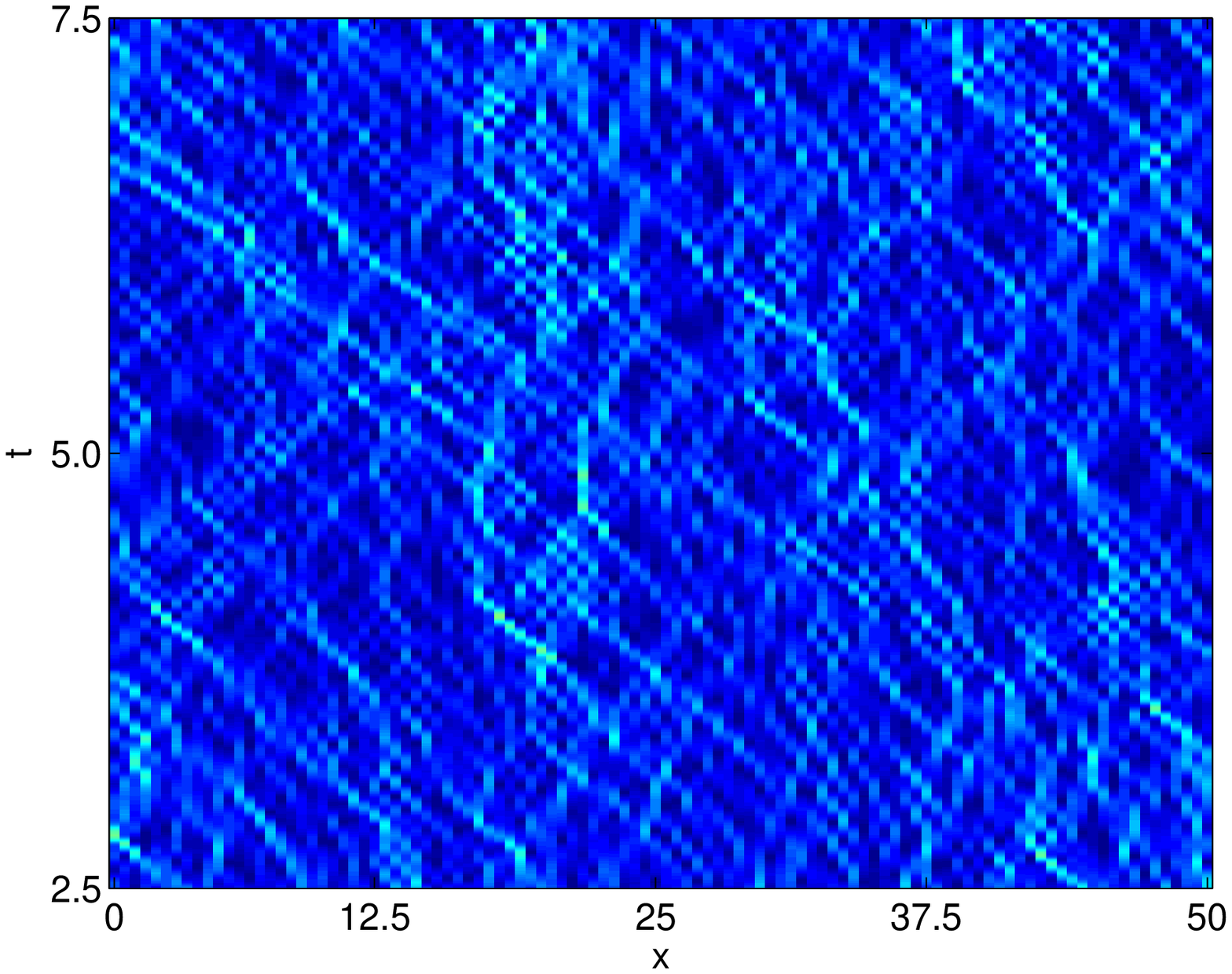}
  \end{center}
\caption{\small (Color online) Skewed chaotic wave-current field of the KEE for $\beta=0.67$.}
  \label{fig2}
\end{figure}

\begin{figure}[htb!]
\begin{center}
   \includegraphics[width=3.4in]{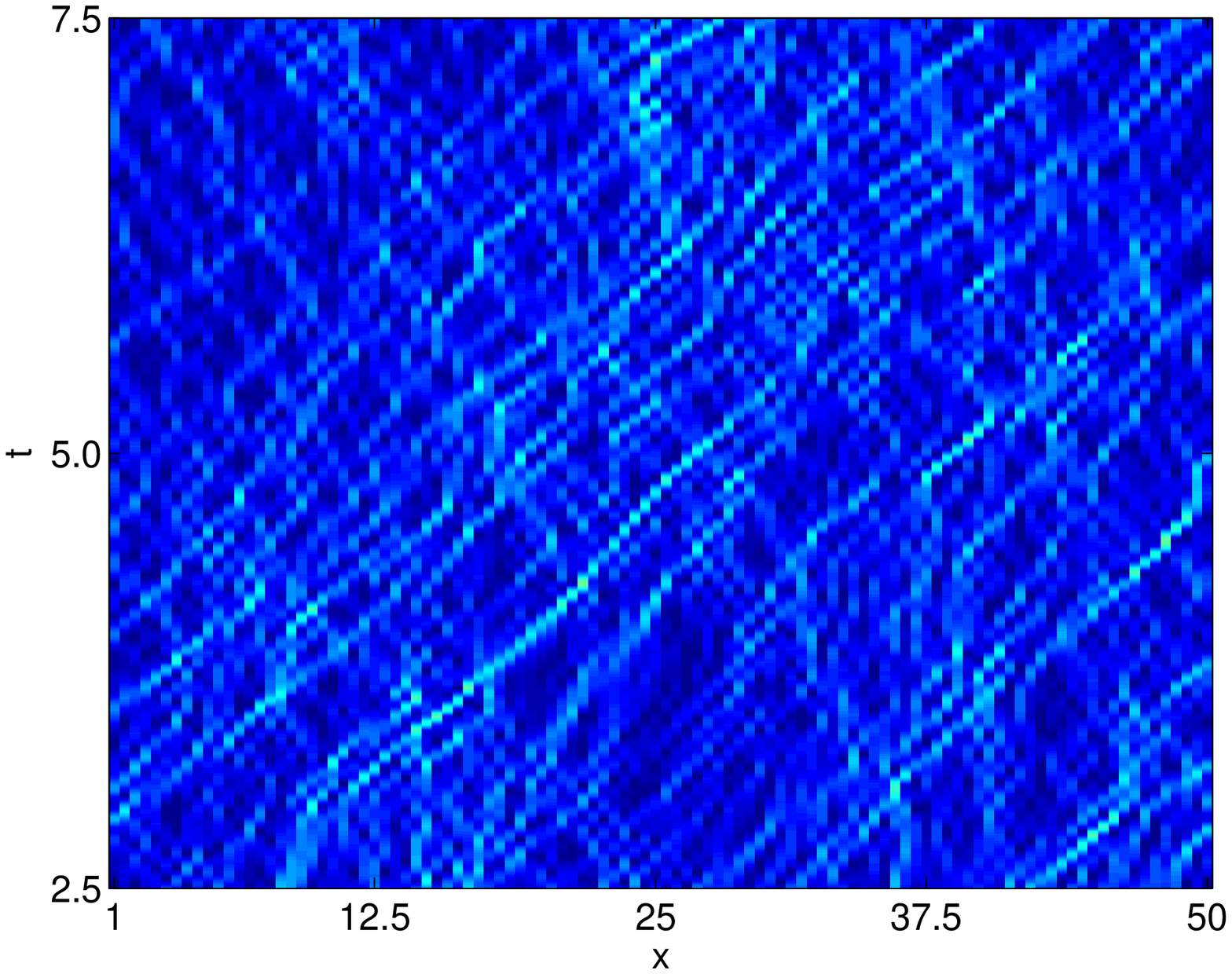}
  \end{center}
\caption{\small (Color online) Skewed chaotic wave-current field of the KEE for $\beta=-0.67$.}
  \label{fig3}
\end{figure}

In order to illustrate the skewed shape of the chaotic wave-current field a contour plot of the $x,t$ plane is shown Fig.\ref{fig2} and Fig.\ref{fig3}. As also described in \cite{bayindir2016KEE, Wang}, the sign of the $\beta$ parameter controls the sign of the skew angle which is oriented relative to the ridge of the rogue wave. If $\beta = 0$ then there is no skew angle and the rogue wave solution of the KEE (obtained from the extended KEE by means of the transformation given in Eq.\,(\ref{eq05})) is no different than the Peregrine soliton solution of the NLSE. For $\beta > 0$ the skew angle is in the counter clockwise direction. For $\beta < 0$ the skew angle the clockwise direction \cite{Wang}. Additionally larger the $\beta$ gets so does the skew angle \cite{Wang}. A more detailed analysis of the contributions of different terms of the KEE can be seen in \cite{bayindir2016KEE}. These results confirm that the extended KEE proposed can explain the skewed shape of the rogue waves which appear in a chaotic wave-current field.

\section{\label{sec:level1}Statistics of Big Waves in Chaotic Wave-Current Field}
The probability density functions (PDFs) of amplitudes ($\left| \psi \right|$) in the chaotic field provide important information about the wave field and about the rogue waves in particular. Therefore we obtain the PDFs for various scenarios. We numerically solve the KEE and simulate the chaotic wave-current field within a spatial domain of $[-100,100]$. We have discarded the initial stages of modulation instability in all runs since deviations from unit amplitude is not significant in those stages. In order to get statistically convergent results, we have used long temporal and spatial intervals.  We have divided the range of amplitudes, $\left| \psi \right|$, into $200$ bins in order to obtain relatively smooth PDFs. First we counted the number of maxima appearing in each bin and then by normalization we have obtained the corresponding PDFs.  Each simulation was repeated 50 times with different initial conditions that correspond to the same values of $\mu$ and $\alpha$. For each of the PDF plots, the data we have analyzed includes approximately one million amplitude maxima which allows us to obtain relatively convergent results and smooth PDFs.

\begin{figure}[htb!]
\begin{center}
   \includegraphics[width=3.4in]{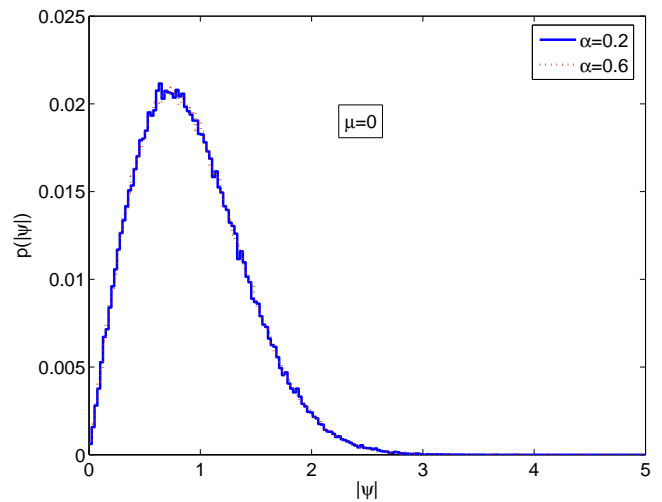}
  \end{center}
\caption{\small (Color online) PDF of the KEE for $\alpha=0.2$ ({\color{blue} ---})  vs. for $\alpha=0.6$ ({\color{red} - . -}) , $\mu=0$.}
  \label{fig4}
\end{figure}

\begin{figure}[htb!]
\begin{center}
   \includegraphics[width=3.4in]{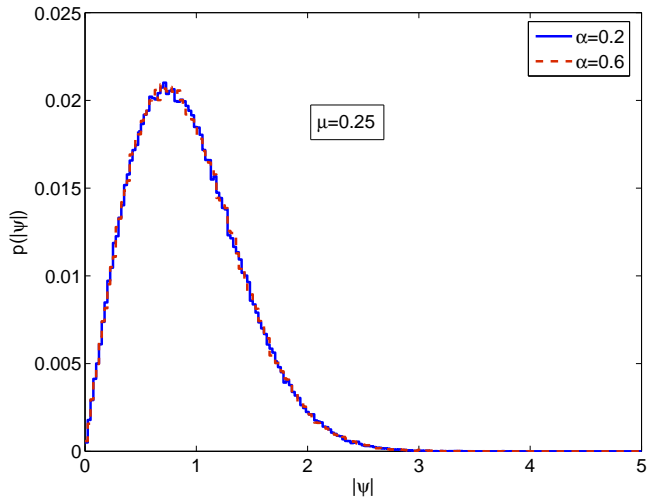}
  \end{center}
\caption{\small (Color online) PDF of the KEE for $\alpha=0.2$ ({\color{blue} ---})  vs. for $\alpha=0.6$ ({\color{red} - . -}) , $\mu=0.25$.}
  \label{fig5}
\end{figure}

\begin{figure}[htb!]
\begin{center}
   \includegraphics[width=3.4in]{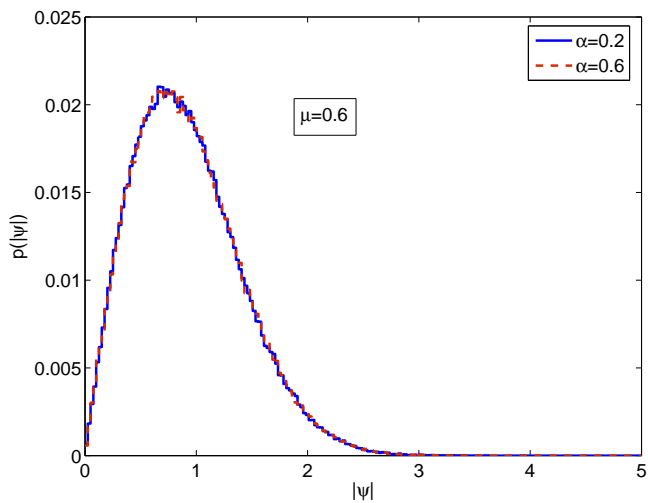}
  \end{center}
\caption{\small (Color online) PDF of the KEE for $\alpha=0.2$ ({\color{blue} ---})  vs. for $\alpha=0.6$ ({\color{red} - . -}) , $\mu=0.6$.}
  \label{fig6}
\end{figure}

In order to quantify the rogue wave occurrence probabilities for different strengths of the chaotic perturbations, we consider three different cases of initial conditions. In the first case of we select a uniform current with $U(x)= -1.57$ and use $\mu=0$, $\alpha=0.2, 0.6$ values respectively for modeling the chaotic current perturbations. In the second case we select the current perturbation parameters as  $U(x)= -4.07$ for $-L \leq x < -10$; $U(x)= -4.07+ \mu x$ for $-10 \leq x \leq 10$ and $U(x)= 0.93$ for $10 < x \leq L$. Setting the parameter $\mu=0.25$, the point $x_o=0$ becomes the blocking point for a plane-wave. We implement chaotic perturbations in the current profile by using $\alpha=0.2, 0.6$. Lastly, we consider a steeper current profile and for this purpose we select parameters as  $U(x)= -7.57$ for $-L \leq x < -10$; $U(x)= -7.57+ \mu x$ for $-10 \leq x \leq 10$ and $U(x)= 4.43$ for $10 < x \leq L$. Selecting  $\mu=0.6$, the point $x_o=0$ becomes the blocking point for a plane-wave . Again using the values of $\alpha=0.2, 0.6$ we perturb the current profile. The PDFs obtained from the numerical simulations for these three cases are shown in the Figs.~\ref{fig4}-~\ref{fig6}. 

One can infer from these results that PDFs closely follow a theoratical Rayleigh distribution. From these results it is clear that the distributions of wave maxima of the chaotic wave-current field, thus the energy level in the chaotic wave field is independent of the parameters $\mu$ and $\alpha$, the mean slope of the current gradient and the magnitude of the chaotic ocean current perturbations defined by Eq. (\ref{eq10}). This is expected because the last term in the extended KEE given in Eq.(\ref{eq04}) introduces phase shifts to the solutions of the KEE, it does not amplify the energy level of the chaotic wave field of the KEE. The average energy for both of the three cases presented in Figs.~\ref{fig4}-~\ref{fig6} are $E \left[ \left|\psi \right|^2 \right]=1.08$. This result, together with the PDFs confirm that the last (extension) term of KEE does not cause an increase in the energy level of the chaotic wave-current field. It rather behaves like a phase shift term, altering the location of occurrence of waves, thus the rogue waves. This can also be verified analytically, the transformation given by Eq.(\ref{eq05}) only introduces a shift to the solutions of the KEE, thus to the rogue waves.

\section{\label{sec:level1}Conclusion}
In this paper we have proposed an extended Kundu-Eckhaus equation for modeling the dynamics of skewed rogue waves emerging in the vicinity of a wave blocking point due to opposing current. The equation we have proposed is a Kundu-Eckhaus equation with an additional potential term therefore the results presented in this paper can easily be generalized to study the quantum tunneling properties of the rogue waves and ultrashort (femtosecond) pulses modeled in the frame of Kundu-Eckhaus equation. In the frame of the extended Kundu-Eckhaus equation, we have numerically showed that the chaotic perturbations of the ocean current trigger the occurrence of the rogue waves on the ocean surface.  We have proposed and implemented a split-step scheme for the numerical solution of the proposed extended Kundu-Eckhaus equation. We have showed that the extended Kundu-Eckhaus equation is unstable against random chaotic perturbations in the opposing ocean current profile. These perturbations transform the initially monochromatic wave field into a chaotic sea state with many peaks. We have numerically showed that the shape of rogue waves due to perturbations in the current profile closely follow the form of rational rogue wave solution of the Kundu-Eckhaus equation, especially for the central peak where the discrepancy is less due to surrounding random smaller waves. We have also discussed the effects of magnitude of the chaotic current perturbations on the statistics of the rogue wave occurrence and showed that stronger perturbations of the ocean current do not generate higher waves on the ocean surface in the frame of the extended extended Kundu-Eckhaus since the extension term of the extended Kundu-Eckhaus equation alters the phases of the solutions of the Kundu-Eckhaus, not their amplitude nor the energy level in the chaotic wave-current field.


\begin{thebibliography}{00}

\bibitem{Akhmediev2011}
N. Akhmediev, J. M. Soto-Crespo, A. Ankiewicz and N. Devine. Physics Letters A,  {\bf{375}}, 2999 (2011).

\bibitem{bayindir2016}
C. Bay\i nd\i r. Physics Letters A,  {\bf{380}}, 156 (2016).

\bibitem{Wang}
X. Wang, B. Yang, Y. Chen and Y. Yang. Physica Scripta,  {\bf{89}}, 095210 (2014).

\bibitem{Akhmediev2009b}
N. Akhmediev, A. Ankiewicz and J. M. Soto-Crespo. Physical Review E,  {\bf{80}}, 026601 (2009).

\bibitem{Kharif}
C. Kharif and E. Pelinovsky. European Journal of Mechanics B: Fluids, {\bf{6}}, 603 (2012).

\bibitem{Smith1976}
R. Smith. Journal of Fluid Mechanics,  {\bf{77}}, 417 (1976).

\bibitem{PeregrineSmith}
D. H. Peregrine and R. Smith. Philosophical Transactions of the Royal Society of London A,  {\bf{292}}, 341 (1979).

\bibitem{dqiu}
D. Qiu,  J. He, Y. Zhang and K. Porsezian. Proceedings of the Royal Society A,  {\bf{471}}, 20150236 (2015).


\bibitem{Akhmediev2014}
J. M. Soto-Crespo, N. Devine, N. P. Hoffmann and N. Akhmediev. Physical Review E,  {\bf{90}}, 032902 (2014).

\bibitem{Akhmediev2015rwsSS}
N. Akhmediev, J. M. Soto-Crespo, N. Devine and N.P. Hoffmann. Physica D,  {\bf{294}}, 37 (2015).


\bibitem{bayindir2016KEE}
C. Bay\i nd\i r. arXiv Preprint,  arXiv:1601.00209 (2016).

\bibitem{Akhmediev2009a}
N. Akhmediev, J. M. Soto-Crespo and A. Ankiewicz. Physics Letters A,  {\bf{373}}, 2137 (2009).


\bibitem{bayindir2009}
C. Bay\i nd\i r. MS Thesis, University of Delaware (2009).

\bibitem{Karjadi2010}
E. A. Karjadi, M. Badiey and J. T. Kirby. The Journal of the Acoustical Society of America,  {\bf{127}}, 1787 (2010).

\bibitem{Karjadi2012}
E. A. Karjadi, M. Badiey, J. T. Kirby and C. Bay\i nd\i r. IEEE Journal of Oceanic Engineering,  {\bf{37-1}}, 112 (2012).


\bibitem{bayindir2016earlyCS}
C. Bay\i nd\i r. arXiv Preprint,  arXiv:1602.00816 (2016).

\bibitem{bay2015d}  C. Bay\i nd\i r. Okyanus dalgalar\i n\i n s\i k\i \c{s}t\i r\i labilir Fourier tayf\i \ y\"{o}ntemiyle h\i zl\i  \ modellenmesi, XIX. T\"{u}rk Mekanik Kongresi, Trabzon, (2015). (In Turkish)

\bibitem{bay2015e}  C. Bay\i nd\i r. S\"{o}n\"{u}ml\"{u} de\u{g}i\c{s}tirilmi\c{s} Korteweg de-Vries (KdV) denkleminin analitik ve hesaplamal\i \ \c{c}\"{o}z\"{u}m kar\c{s}\i la\c{s}t\i rmas\i, T\"{u}rk Mekanik Kongresi, Trabzon, (2015). (In Turkish)

\bibitem{trefethen}
L. N. Trefethen. Spectral Methods in {MATLAB}, (2000).

\bibitem{bayindir2016nature}
C. Bay\i nd\i r. Sci. Rep.,  {\bf{6}}, 22100; doi: 10.1038/srep22100 (2016).

\bibitem{demiray}
H. Demiray and C. Bay\i nd\i r. Physics of Plasmas,  {\bf{22}}, 092105 (2015).

\bibitem{bayindir2015d}
C. Bay\i nd\i r. TWMS: Journal of Applied and Engineering Mathematics,  {\bf{5-2}}, 298 (2015).


\bibitem{bayindir2015arxivchbloc}
C. Bay\i nd\i r. arXiv Preprint,  arXiv:1512.03584 (2015).

\bibitem{bayindir2015arxivcssfm}
C. Bay\i nd\i r. arXiv Preprint,  arXiv:1512.03932 (2015).


\bibitem{bay2015c}  C. Bay\i nd\i r. Hesaplamal\i \ ak\i\c{s}kanlar mekani\u{g}i  \c{c}al\i\c{s}malar\i \ i\c{c}in s\i k\i\c{s}t\i r\i labilir Fourier tayf\i \ y\"{o}ntemi, XIX. T\"{u}rk Mekanik Kongresi, Trabzon, (2015). (In Turkish)


\bibitem{bayindir2015arxivcsmww}
C. Bay\i nd\i r. arXiv Preprint,  arXiv:1512.06286 (2015).


\bibitem{Peregrine}
D. H. Peregrine. Journal of Australian Mathematical Society: Series B,  {\bf{25}}, 16 (1983).




 \end{thebibliography}
\end{document}